\documentclass[12pt,a4paper,dvips]{article}

\usepackage{a4p}
\usepackage{cite,mcite}
\usepackage{graphicx}
\usepackage{physics}
\usepackage{l3_title}

\newcommand{\EE}{\rm e^+ e^-}

\newlength{\capindent}
\newlength{\capwidth}
\newlength{\figwidth}
\newcommand{\icaption}[2][!*!,!]{\hspace*{\capindent}%
  \begin{minipage}{\capwidth}
    \ifthenelse{\equal{#1}{!*!,!}}%
      {\caption{#2}}%
      {\caption[#1]{#2}}
  \end{minipage}}
\begin{document}
\bibliographystyle{/l3/paper/biblio/l3style}
\begin{titlepage}
\title{ 
Measurement of the Two Photon Decay Width 
of the Higgs Boson at the TESLA Photon Collider}
\author{Aura Rosca and Klaus M\"onig}
\vspace*{-1.5cm}
\begin{center}
{\it DESY Zeuthen, Platanenallee 6, 15738 Zeuthen, Germany}
\end{center}

%

\begin{abstract}
We report on the accuracy of the measurement of the two photon decay width
of a Higgs boson with the mass of 120 GeV at the TESLA Photon Collider,
assuming a $\gamma \gamma$ integrated luminosity of 80 fb$^{-1}$ in the
hard part of the spectrum. The QCD radiative corrections for the quark
pair background processes are taken into account by a reweighting
procedure. We found that the product $\Gamma (\rm H \to 
\gamma \gamma) \times \rm BR (\rm H \to \rm b \bar{\rm b})$ 
can be measured with the statistical error of 1.8$\%$
in one year of run.

\end{abstract}
%
%
\vspace*{20mm}
\end{titlepage}

\section*{Introduction}

The search for the Higgs boson
to explore the mechanism of symmetry breaking is one of
the most important tasks of the experimental program at the future
colliders.
While it can only be produced in association with another particle at an
$\EE$ collider, the Higgs boson can be produced singly in the s-channel
of the colliding photons at a photon collider.
If the mass of the Higgs boson is light, as indicated by the current
electroweak measurements, then it will be found by the time a $\gamma \gamma$
collider is constructed. 
The aim of such collider will be then a precise measurement of Higgs properties.
Among these the measurement of the two photon decay width of the Higgs boson is
particularly important. The Higgs boson can be produced in $\gamma \gamma
$ collisions
via a one-loop diagram, in which all charged particles whose masses derive from the 
Higgs mechanism  can appear inside the loop. In the Standard Model (SM),
the dominant 
contributions come from the top and W boson loops. A deviation of the
two photon width 
from the SM prediction indicates additional contribution from unknown particles, and 
is a signature of physics beyond the SM. For example, the minimal extension of the SM
predicts the ratio of the two photon width $\Gamma(\rm h \to \gamma \gamma, \rm MSSM)$/
$\Gamma(\rm H \to \gamma \gamma, \rm SM) < $ 1.2  \cite{mssm}
for a Higgs boson with a mass of 120 GeV, assuming a supersymmetry
scale of 1 TeV and the chargino
mass parameters $M$ and $\mu$ of 300 and 100 GeV, respectively.

At a $\gamma \gamma$ collider, the two photon decay width is obtained by measuring the
Higgs formation cross section in the reaction $\gamma \gamma \to \rm H \to \rm X$, where
X is the detected final state. The number of detected events is proportional to the
product $\Gamma(\rm H \to \gamma \gamma)$$\times$BR$(\rm H \to \rm X)$.
Measuring the 
formation cross section will determine this product. An independent measurement of the
branching ratio  BR($\rm H \to \rm X$) at $\EE$ collider 
allows a determination of the $\gamma \gamma$ partial width.

In this paper we study the accuracy of the measurement of the two photon decay width for
a Higgs boson with the mass of 120 GeV. The study is performed for the photon collider 
option of the TESLA Linear Collider operated at centre-of-mass energy of
210 GeV. The assumed data
volume corresponds to a total $\gamma \gamma$
integrated luminosity of 400 fb$^{-1}$ \cite{pc},
which amounts
to about ${\cal L}_{\gamma \gamma}(\sqrt s _{\gamma \gamma} > 80 \rm GeV)$
= 80 fb$^{-1}$
in the hard part of the spectrum.

The feasibility of the measurement of the two photon decay width of the Higgs boson in 
this mass region has also been reported by \cite{notes}.

\section*{Simulation of the signal and background processes}

The cross section for the Higgs boson formation is given by a Breit-Wigner approximation
$$\sigma_{\gamma \gamma \to \rm H}=8 \pi \frac{\Gamma(\rm H \to \gamma \gamma) 
\Gamma_{\rm tot}}{(s _{\gamma \gamma}-M_{\rm H}^{2})^{2}+M^{2}_{\rm H}
\Gamma^{2}_{\rm tot}}(1+\lambda_{1}\lambda_{2}),$$
where $M_{\rm H}$ is the Higgs boson mass, $\Gamma(\rm H \to \gamma \gamma)$ and
$\Gamma_{\rm tot}$ are the two photon and total decay width of the Higgs 
boson, 
$\lambda_{1}$ and $\lambda_{2}$ are the initial photon helicities and
$\sqrt s_{\gamma \gamma}$ is the $\gamma \gamma$ centre-of-mass energy.
The initial photons 
should have equal helicities, so that $J_{\rm z}$ = 0, in order to make a
spin-0 
resonance as it is the case of the Higgs boson. If polarised photon beams
are used, 
the signal cross
section is increased up to a factor of 2.
The experimentally 
observed cross section is obtained by folding this basic cross section with the
$\gamma \gamma$ collider luminosity distribution.

A Higgs boson with the mass of 120 GeV is detected in
the  $\gamma \gamma \to \rm H \to \rm b \bar{\rm b}$ final state.
The event rate is given by the formula:
$$N(\gamma \gamma \to \rm H \to \rm b \bar{\rm b})= 
\frac{d{\cal L}_{\gamma \gamma}}{\rm d \sqrt 
s _{\gamma \gamma}}|_{M_{\rm H}} \frac{4 \pi^{2}\Gamma (\rm H \to \gamma \gamma)
\rm BR (\rm H \to \rm b \bar{\rm b})}{M_{\rm H} ^{2}} (1+\lambda_{1}\lambda_{2})(\hbar c)^{2},$$
where the conversion factor $(\hbar c)^{2}$ is  3.8937966$\cdot$$10^{11}$ fb GeV$^{2}$.
 
\begin{figure}[H]
\begin{center}
\includegraphics[width=10.5cm]{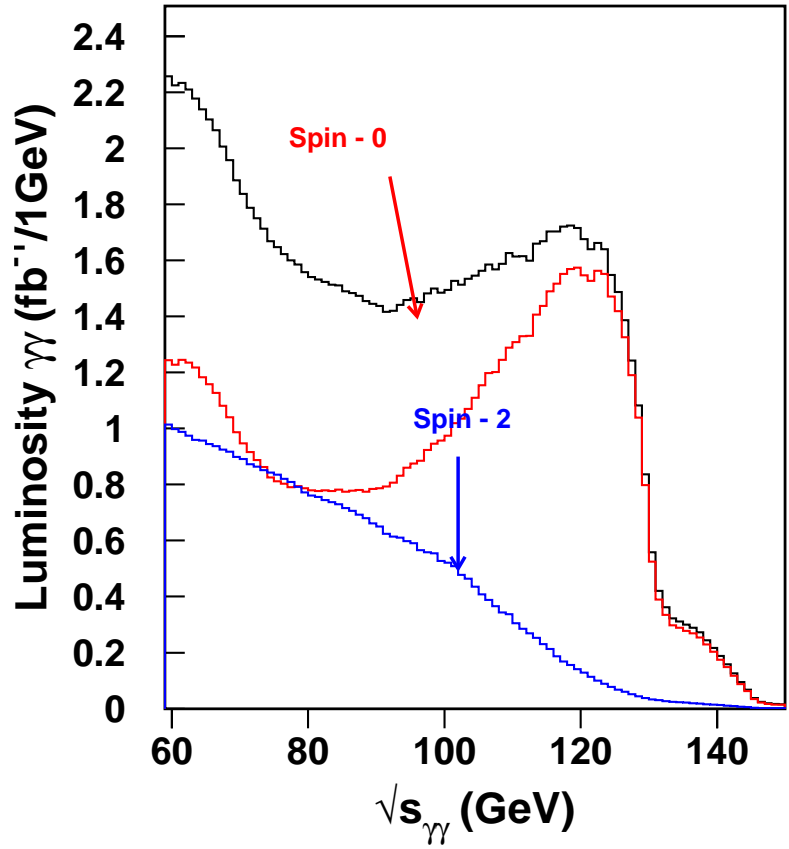} \\
\caption{Luminosity spectra for the spin of the two colliding photons
$J_{z}$ = 0 and $J_{z}$ = 2.
}
\label{fig_lumi}
\end{center}
\end{figure}

It depends strongly on 
the differential luminosity, $\frac{d{\cal L}_{\gamma \gamma}}{\rm d \sqrt 
s _{\gamma \gamma}}|_{M_{\rm H}}$. The shape of the luminosity distribution depends on the
electron and laser beam parameters. The electron and laser beam energy 
considered for this study are 105 GeV and
1 eV, respectively, resulting in the maximum photon energy of about 75 
GeV, suitable 
to study a Higgs boson with the mass of 120 GeV. 
The combination of the polarisation of the laser and the electron are
chosen such to make
the generated photon spectrum peak at its maximum energy. The helicity combination
of the two high energy photons is arranged such that $J_{\rm z}$ = 0 state
is dominant.
The luminosity distributions for the $J_{\rm z}$=0 and $J_{\rm 
z}$ = 2 are provided by the CIRCE2 \cite{circe} program and shown in 
Figure \ref{fig_lumi}. 
Beamstrahlung,
secondary collisions between electrons and laser beam photons, and other non-linear 
effects are taken into account when calculating the $\gamma \gamma$ luminosity.
The resulting value of $\frac{d{\cal L}_{\gamma \gamma}}{\rm d \sqrt 
s _{\gamma \gamma}}|_{M_{\rm H}}$ is 1.6 fb$^{-1}$/GeV. 

The branching ratios BR(H $\to \gamma \gamma$), BR(H $\to \rm b \bar{\rm b}$) and 
the total width are taken to be 0.22$\%$, 68$\%$ and 4 MeV, respectively, and are
calculated with the HDECAY \cite{hdecay} program including QCD radiative corrections.
A signal rate of about 20000 events per year can be achieved under these 
conditions.

The main background processes are the
direct continuum $\gamma \gamma \to \rm b \bar{\rm b}$ and
$\gamma \gamma \to \rm c \bar{\rm c}$ production. 
Due to helicity conservation, the
continuum background production proceeds mainly through the states of opposite photon 
helicities, making the states $J_{\rm z} = 2$. Choosing equal helicity photon 
polarisations the 
cross section of the continuum background is suppressed by a factor 
$M_{\rm q}^{2}/s_{\gamma \gamma}$, with $M_{\rm q}$
being the quark mass. Unfortunately, this suppression does not apply
to the process $\gamma \gamma \to \rm q \bar{\rm q} \rm g$, because after the gluon 
radiation the $\rm q \bar{\rm q}$ system is not necessarily in a
$J_{\rm z} = 0$ 
state. The surviving background is large and overwhelms the signal. 

\begin{figure}[H]
\begin{center}
\includegraphics[width=10.5cm]{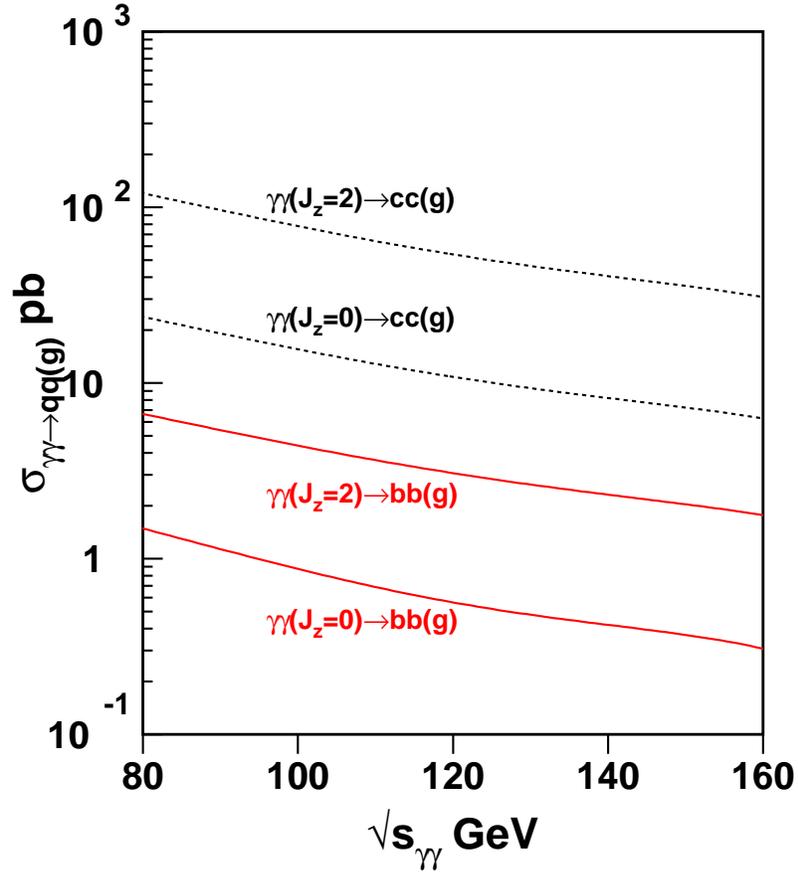} \\
\caption{Cross sections for the $\gamma \gamma \to \rm b \bar{\rm b} (\rm g)$
and $\gamma \gamma \to \rm c \bar{\rm c} (\rm g)$ processes including QCD radiative
corrections for $J_{z}$ = 0 and $J_{z}$ = 2.
}
\label{fig_cs}
\end{center}
\end{figure}

The signal $\gamma \gamma 
\to {\rm H} \to {\rm b} \bar{\rm b}$ process, as well as the background
$\gamma \gamma \to {\rm q} \bar{\rm q}$(g) events are generated with the PYTHIA 
\cite{pythia} program.
Parton evolution and hadronization are simulated using
the parton shower and the string fragmentation models.
A convolution with the luminosity distribution is performed, and kinematic
cuts of $|\cos \theta|<0.9$ and $\sqrt s_{\gamma \gamma}$ greater than 80 GeV 
are imposed during the event generation, $\theta$ denoting the polar angle of the 
produced quarks. 
The QCD corrections to $\gamma \gamma \to {\rm q} \bar{\rm q}$ have to be taken into 
account for a reliable background estimation. 
For this purpose we use the total cross
sections 
shown in Figure \ref{fig_cs}, which include the soft and hard gluon
emission \cite{qcd1}, virtual corrections \cite{qcd1} and non-Sudakov form factors 
\cite{qcd2} to weight the event. 
The event topology is given by the parton shower model. 
 
The effective cross sections, the
number of expected and generated signal and background events are 
presented in Table \ref{tab_cs}. 

To simulate the response of the apparatus
we use SIMDET \cite{simdet}, a fast Monte Carlo program which smears 
the kinematics of the particles in the final state according to the
detector resolutions. The detector used in the simulation follows the proposal
presented in the Technical Design Report of the TESLA $\rm e^{+} \rm e^{-}$ 
Linear Collider \cite{tdr}. 
However, the changes in the detector design for $\gamma \gamma$ collisions will affect 
only the forward region \cite{ggdet} which is not important for this analysis.

\section*{Event selection}

The analysis aims to select events with two or three multiplicity jets 
from the Higgs boson decay. Two of these jets contain bottom quarks. The invariant
mass of the jets has to be consistent with the Higgs mass.

Jets are reconstructed using the DURHAM clustering scheme \cite{durham} 
with the parameter y$_{\rm cut}$=0.02. 
 
\begin{figure}[H]
\begin{center}
\includegraphics[width=10.5cm]{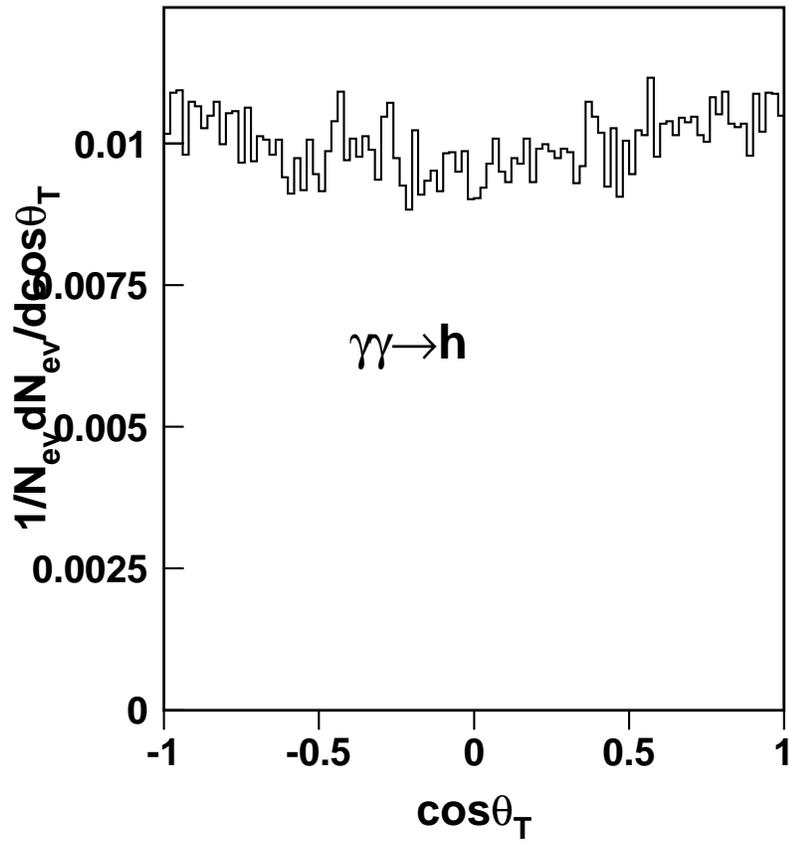}
\includegraphics[width=10.5cm]{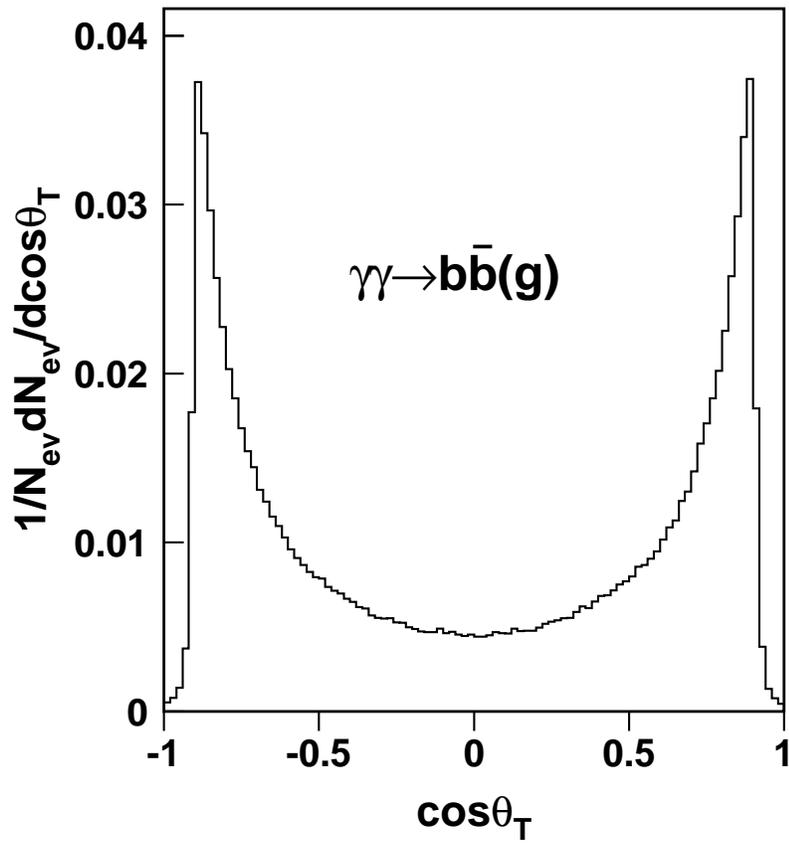} \\
\caption{Distributions for the cosine of the thrust angle  
for signal a) and background b) events.
}
\label{fig_costh}
\end{center}
\end{figure}

We select events with a visible energy larger than 95 
GeV and the energy imbalance longitudinal to the beam direction below 10$\%$
of the visible energy. 
Figures \ref{fig_costh}a,b shows the distribution of the cosine of the thrust angle 
for the signal
and background events. The s-channel signal process has an isotropic angular distribution,
while the t-channel background processes are forward peaked.
We require the absolute value of this quantity to be below 0.7. 

\begin{figure}[H]
\begin{center}
\includegraphics[width=10.5cm]{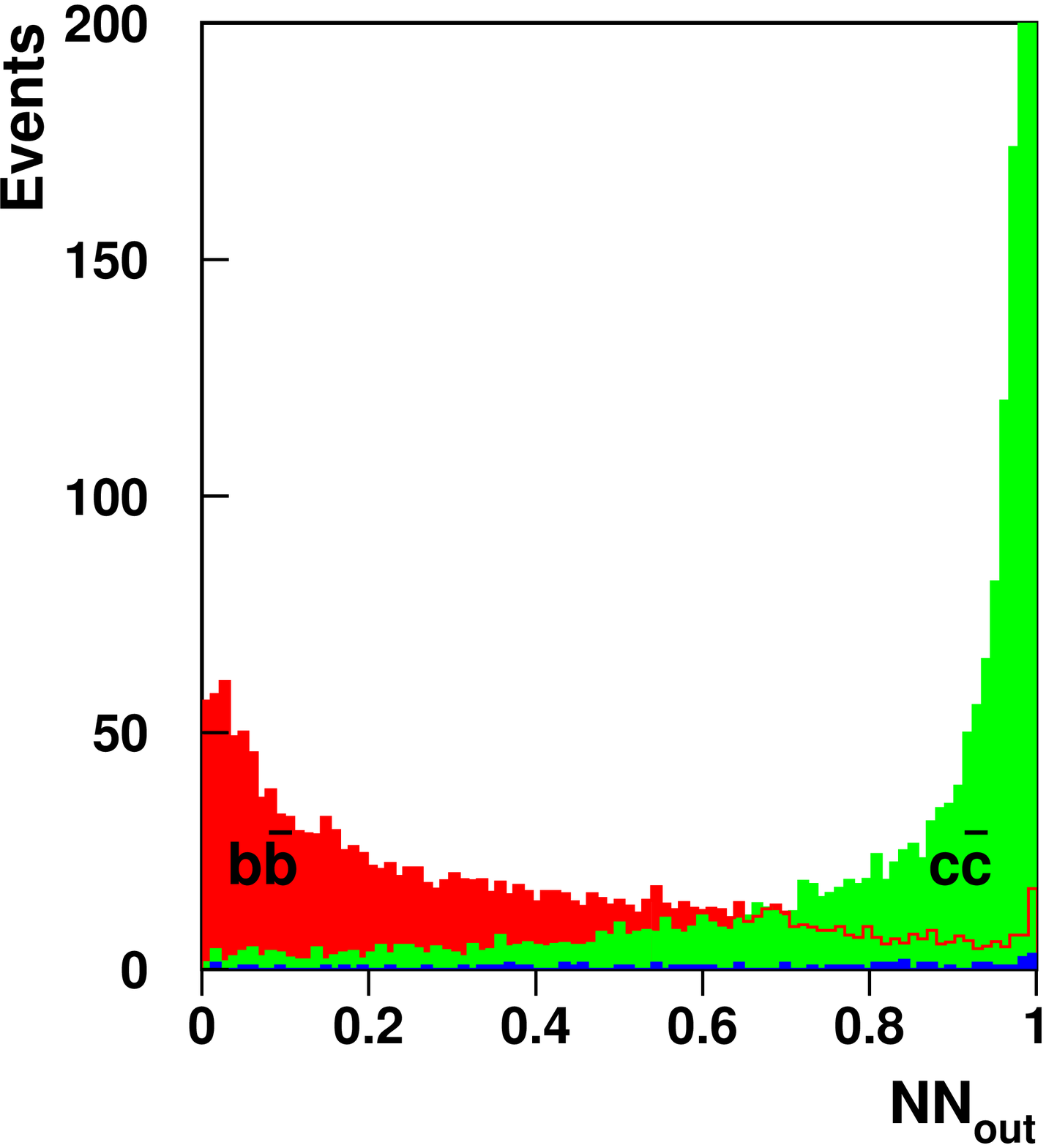}
\includegraphics[width=10.5cm]{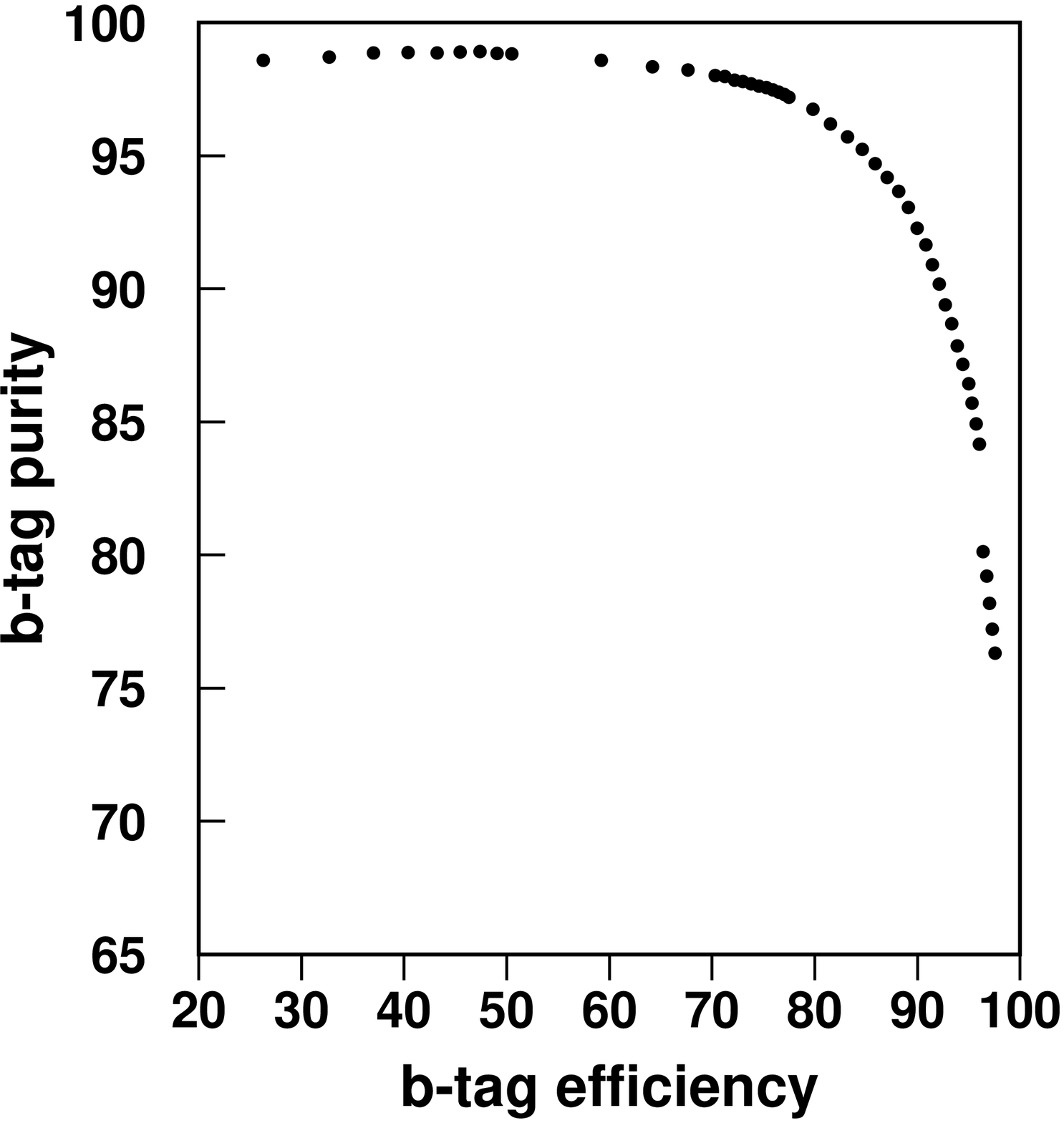} \\
\caption{a) Distribution of the neural network output and 
b) efficiency versus purity curve for the neural network based b-tag in Z$^{0}$ decays.
}
\label{fig_btag}
\end{center}
\end{figure}

The cross section for the continuum production of the charm quark is 16 times
larger than for bottom quarks, therefore b-quark tagging 
is crucial in this analysis. The b-tagging 
algorithm relies on a neural network  with 12 inputs and 3 output nodes, described 
in Ref. \cite{nn}. The most important inputs comprise the impact parameter 
joint probability tag 
introduced by ALEPH \cite{aleph}, the $p_{t}$ corrected vertex invariant mass obtained with
the ZVTOP algorithm written for the SLD experiment \cite{zvtop} and a one-prong charm tag using the
largest and second largest track impact parameter significances in $r$ - $\phi$
and $r$ - $z$, in jets where ZVTOP found only the primary vertex.   
The spectrum of the neural network output $NN_{\rm out}$ to discriminate
b-quark jets
from u-, d-, s- and c-quark jets 
is presented in Figure \ref{fig_btag}a. The performance of the neural network b-tag in Z$^{0}$
decays is shown in Figure \ref{fig_btag}b.  
We require at least 2 vertices to be found and ask that the $NN_{\rm out}$
to be greater 
than 0.95. The b-tagging efficiency is 70$\%$ and corresponds to a purity of 98$\%$.

The efficiency of this selection is found to be 36$\%$.
 
\section*{Discussion of the QCD {\boldmath $\gamma \gamma \to {\rm
hadrons}$} background}

A background not discussed yet is the process $\gamma \gamma \to$hadrons,
where photons interact hadronically at the interaction point.
With a total hadronic cross section in $\gamma \gamma$ collisions of 400 nb, 
we estimate about 1.4 such events per
bunch crossing. 
These events obscure the interesting physics processes described in the
previous sections. For this reason this class of events needs to be 
included in the PYTHIA simulation for overlap.
The HADES \cite{hades} program is used for this purpose.
A large fraction of this background is distributed at small angles and we
believe that it can be reduced cutting on the polar angle of the tracks
\cite{aura}. A careful study of this hadronic background is currently
being performed.

\section*{Results}

\begin{figure}[H]
\begin{center}
\includegraphics[width=10.5cm]{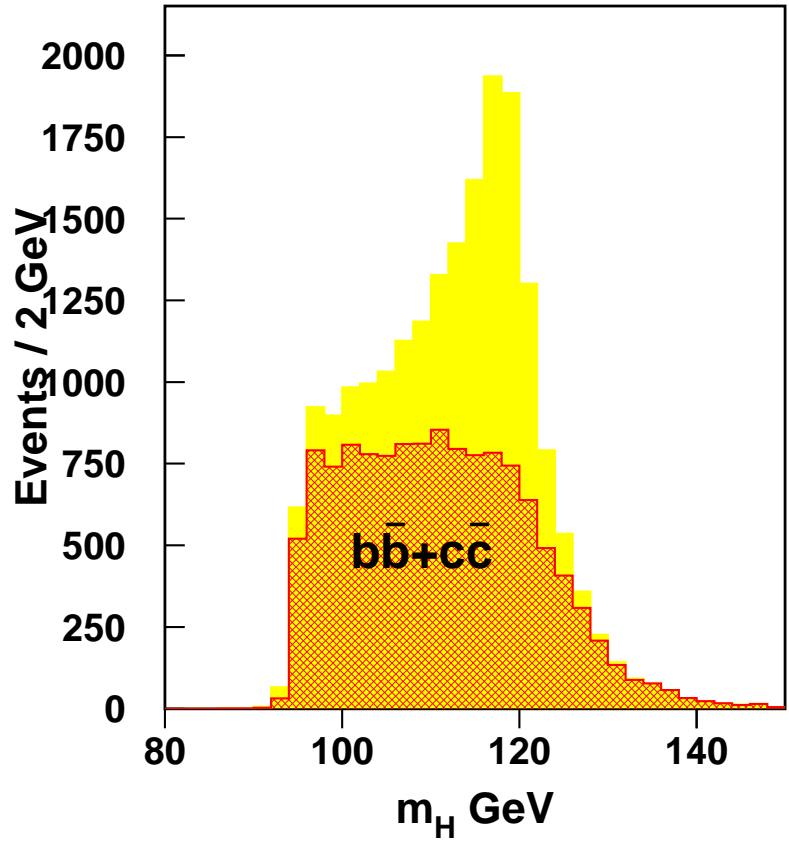} \\
\caption{Distribution of the reconstructed invariant mass  
for the signal and background events.
}
\label{fig_mass}
\end{center}
\end{figure}

The reconstructed invariant mass for the selected signal and background
events 
is shown in Figure \ref{fig_mass}.  
To enhance the signal a cut on the invariant mass is 
tuned such that the statistical significance of the signal over background 
is maximised. Events in the mass region of 106 GeV $ < M_{jj} < $ 126 GeV
are selected. The number of estimated signal and background events 
in this window are 6018 and 7111, respectively.

The two photon decay width of the Higgs boson is
proportional to the event
rates of the Higgs signal. The statistical error of the number of signal events,
$\sqrt N_{\rm obs}/({\it N} _{\rm obs}-{\it N}_{\rm bkg})$,
corresponds to the statistical error of this measurement. Here $N_{\rm obs}$ is the
number of observed events, while $N_{\rm bkg}$ is the number of expected background
events. 

We obtain
$$\frac{\Delta [\Gamma (\rm H \to \gamma \gamma)\times \rm BR (\rm H \to \rm b
\bar{\rm b})]}{[\Gamma (\rm H \to \gamma \gamma) \times \rm BR (\rm H \to \rm b
\bar{\rm b})]}=\sqrt N_{\rm obs}/({\it N} _{\rm obs}-{\it N}_{\rm bkg})=1.9\% .$$
We can improve this number by 0.1$\%$ doing a fit with the function
$$ \chi^{2}=\Sigma \frac{((f_{i}N_{\rm H}+b_{i})^{2}-d_{i}^{2})^{2}}{d_{i}},$$
with $f_{i}=\frac{s_{i}}{s}$ being the fraction of signal events, $b_{i}$ and $d_{i}$ the
background and observed number of events in each bin of the invariant mass
distributins and $N_{\rm H}$ a fit parameter.

We conclude that for a Higgs boson with a mass $M_{\rm H}$=120 GeV 
we can measure the product $\Gamma (\rm H \to 
\gamma \gamma) \times \rm BR (\rm H \to \rm b \bar{\rm b})$ 
with an accuracy of 1.8$\%$ using an integrated luminosity corresponding to
one year of data taking at the TESLA Photon Collider.


\newpage
\begin{table} [ht]
\begin{center}
\hspace*{-1.cm}
\begin{tabular}{|c|c|c|c|}\hline
                  &Cross section
& Number of & Number of\\ 
                  &(pb)&expected events&generated events\\ 
\hline
Signal process & $\sigma_{\rm eff}$, ${\cal L}$ = 80 fb$^{-1}$& 
~ 20000& 50000 \\
$\gamma \gamma \to \rm H \to {\rm b} \bar{\rm b}$
& 0.25&  & 
 \\
\hline
Background &    & & \\
$\gamma \gamma \to {\rm b} \bar{\rm b}$(g) 
&0.75 & 44175.0& 600000 \\
$J$ = 0 & & (${\cal L}_{J=0}=58.9$ fb$^{-1}$)& \\
\hline
$\gamma \gamma \to {\rm b} \bar{\rm b}$(g) 
&4.79 & 102314.4& 600000 \\
$J$ = 2 & & (${\cal L}_{J=2}=21.1$ fb$^{-1}$)& \\
\hline
$\gamma \gamma \to {\rm c} \bar{\rm c}$(g) 
&13.4 &789260.0& 600000 \\
$J$ = 0 & & & \\
\hline
$\gamma \gamma \to {\rm c} \bar{\rm c}$(g) 
&85.1 &1817734.0& 600000 \\
$J$ = 2 & & & \\
\hline
\end{tabular}
\caption{
Cross sections and the number of expected and generated events for the 
signal and 
background processes. 
}
\label{tab_cs}
\end{center}
\end{table}

%
\newpage
\newpage
\newpage
\newpage
\newpage

\end{document}